\documentstyle[12pt]{article}
\begin{document}
\rightline{{\bf YCTP-P38-92}}
\rightline{September 1992}
\baselineskip=18pt
\vskip 0.2in
\begin{center}
{\bf\large SOLAR NEUTRINO DATA, SOLAR MODEL UNCERTAINTIES
  AND NEUTRINO OSCILLATIONS}
\end{center}
\vskip 0.1in
\begin{center}
Lawrence Krauss\footnote{\noindent Also Department of Astronomy.
Research supported in part by the DOE, the NSF and the TNRLC. Bitnet:
KRAUSS@YALEHEP}

{\it Center for Theoretical Physics}

{\it Sloane Laboratory \it Yale University}

{\it New Haven, CT 06511, USA}

\vskip .2in
Evalyn Gates\footnote{\noindent Supported in part by the DOE.}

{\it Astronomy and Astrophysics Center}

{\it Enrico Fermi Institute, University of Chicago}

{\it Chicago IL 60637}

\vskip .2in
Martin White

{\it Center for Particle Astrophysics}

{\it University of California, Berkeley, CA 94720}
\end{center}

\vskip .5in
\centerline{ {\bf Abstract} }

\noindent
We incorporate all existing solar neutrino flux measurements and take solar
model flux uncertainties into account in deriving global fits to parameter
space for the MSW and vacuum solutions of the solar neutrino problem.

\newpage

The solar neutrino problem continues unabated after over 20 years and
four experiments.  In spite of the relatively
high flux recently reported by the GALLEX \cite{GALLEX} collaboration,
a neutrino based solution of the solar neutrino problem still remains
strongly favored \cite{White}, at least as long as the Homestake Cl
\cite{Homestake} result remains unchanged. (Note that the 20 year Homestake
weighted average used here and in most analyses is lower than an
unweighted average, or an average over only the last 3 year's data.  If a
higher
Homestake value is used, this can affect the comparative range of various
neutrino-based solutions.)  In this paper, we investigate what range of
neutrino masses and mixing angles (for 2 species mixing) remains consistent
with a global fit of {\it all} the data, including the updated 2 year SAGE
average of $58^{+17}_{-24} \pm 14$ SNU \cite{SAGE}.
We stress the following: (1) The MSW range is not the only range of masses and
mixing angles which is consistent with the observations.  Indeed, while the
original MSW parameter space allowed by the Homestake data was far larger than
for vacuum oscillations, this difference is now much less significant, at least
for the canonical small angle MSW region; (2)  There is no justification for
incorporating the new GALLEX result and not the SAGE result in model fits; (3)
Solar model uncertainties noticeably increase the allowed range of neutrino
parameter space which is consistent with the existing observations.

Neutrino based solutions of the solar neutrino problem fall into 2
major classes, related to suppression of the neutrino signal at the
earth due to either oscillations between neutrino flavors or
helicity states.  Both involve at least one non-zero neutrino mass.
The latter type involves potentially non-trivial time dependence of the
solar neutrino signal over the solar cycle, but requires extremely
large neutrino magnetic moments to remain viable, and has been recently
investigated elsewhere \cite{Gates}.  The former falls into two
sub-categories: resonant neutrino oscillations inside the sun due to
a changing electron density in the solar core (MSW oscillations \cite{MSW}),
and oscillations in vacuo between the earth and sun (vacuum
oscillations).  The latter of these involves masses which are 2-3
orders of magnitude smaller, and consequently vacuum oscillation
lengths which are up to 6 orders of magnitude larger than those
appropriate for MSW matter oscillations.

Both MSW and vacuum oscillations result in an energy dependent suppression of
the neutrino signal.  The latter has a clear oscillation in energy
\cite{GlaKra}.   In principle, this could be useful to distinguish between
these possibilities once high statistics measurements of the solar neutrino
spectrum are available.  We find that the Kamiokande
\cite{Kamiokande,Kamiokande2} spectral measurements do not presently constrain
regions of MSW parameter space beyond those which can be ruled out out by
global
fits to overall flux measurements. Vacuum oscillations also allow a possible
seasonal variation in the observed signal, as the earth-sun distance changes,
and different oscillation lengths are sampled.
Time sequenced data is available for each of the experiments.  However, there
is a great deal of scatter in the data, and it is not clear at this point to
what extent any systematic time variation can be extracted from the signal (
except that a systematic day-night variation which would occur for certain of
the MSW parameter space is not observed \cite{Kamiokande3}). Thus, in this
paper, we merely compare the predicted time-averaged total signals in the
Homestake, Kamiokande, SAGE and GALLEX experiments with the observed signals
in order to derive, by statistical methods, the region of parameter space which
remains viable, {\it{after}} solar model uncertainties have been taken into
account.  Additional tests based on spectral information, or time dependence
can serve to further reduce this region, but probably only marginally so at
present.   Several recent analyses have examined these latter issues in more
detail, especially for vacuum oscillations \cite{Petcov,Barger}.

The strategy for calculating the neutrino survival probability at the earth
is different in the case of vacuum oscillations and MSW oscillations.  In
both cases, we consider here only possible oscillations to an active
neutrino, as this remains at present the most likely possibility now that
evidence for a 17 keV neutrino has diminished \cite{17keV}.
For the case of vacuum oscillations \cite{GlaKra} the details of the SSM
production regions and $N_e$ density profile are unimportant, the $\nu_e$
survival probability $P_{vac}$ at a distance $L$ from the center of the sun
is simply \cite{GlaKra,TheBook}
\begin{equation}
  P_{vac}(\nu_e \rightarrow \nu_e)
    = 1 - \sin^2 2\theta \sin^2\left( {\Delta m^2 L\over 4E} \right)
\end{equation}
We averaged $P_{vac}$ over a varying Earth-Sun distance to take into account
(in a simple minded way) the motion of the Earth around the Sun during each
measurement.
The average was taken (linearly) from $L(1-\epsilon)$ to $L(1+\epsilon)$ with
$\epsilon$ equal to the Earth's orbital eccentricity \cite{Gates}.

For the MSW model \cite{MSW,TheBook} there exist good analytic approximations
to the $\nu_e$ survival probability under the assumption that the electron
number density ($N_e$) decreases exponentially $N_e\sim\exp(-r/r_0)$
near the resonance.
We used the approximations and best fit values of $r_0$ given in \cite{KP} who
break the $(\Delta m^2,\sin^22\theta)$ into several regions, giving
approximations good to a few percent in each region.
A natural distinction is between adiabatic and non-adiabatic transitions.
If we define:
\begin{equation}
  4n_0 = r_0 \left( {\Delta m^2 \over {2E}} \right)
  \left( {\sin^2 2\theta \over\cos 2\theta} \right)
\label{eqn:n0}
\end{equation}
then the non-adiabatic region is given by the condition $4n_0\leq 1$
while in the adiabatic region $4n_0\gg 1$ (a reasonable choice for the
transition value is $4n_0= 4$) \cite{Petcov}.
We then subdivide the non-adiabatic region into three parts, depending on the
size of $N_e^{res}=\Delta m^2\cos 2\theta/2\sqrt{2}G_FE$ compared to
$N_e^{(1)} = N_e^{(0)}(1 + \tan 2\theta)^{-1}$ and
$N_e^{(2)} = N_e^{(0)}(1 - \tan 2\theta)^{-1}$, with $N_e^{(0)}$ the
electron density where the $\nu$ is produced.  Writing
\begin{eqnarray}
             x & = & 2\pi\ r_0{\Delta m^2 \over 2E} \\
             y & = & 2\pi\ n_0(1 - \tan^2 \theta) \\
\cos 2\theta_m & = & (1 - \eta) / \sqrt{ (1-\eta)^2+\tan^2 2\theta} \ \ \
   \eta\equiv N_e^{(0)} / N_e^{res}
\end{eqnarray}
the analytic expressions for the $\nu_e$ survival probability we used were

\vspace{0.1in}

\noindent\underline{$N_e^{res} < N_e^{(1)}$}: \newline
\begin{equation}
\bar{P}(\nu_e\rightarrow\nu_e) =
  {1\over2} + \left( { 1 + e^{-x} \over 1 - e^{-x} } \right)
  \left[ {1\over 2} - {e^{-y}\over 1 + e^{-x}}\right]
  \cos 2\theta_m \cos 2\theta,
\label{eqn:PNA}
\end{equation}

\noindent\underline{$N_e^{(1}) \leq N_e^{res} \leq N_e^{(2)}$}: \newline
\begin{equation}
\bar{P}(\nu_e \rightarrow \nu_e) =
  {1\over 2} \left[ 1 + \exp (-\pi n_0)\right].
\label{eqn:P0NA}
\end{equation}

\noindent\underline{$N_e^{res} > N_e^{(2)}$}: \newline
\begin{equation}
\bar{P}(\nu_e\rightarrow\nu_e)
  = {1\over 2} + {1\over 2}\cos 2\theta_m \cos 2\theta.
\label{eqn:PA}
\end{equation}
For neutrinos in the adiabatic region we used equation (\ref{eqn:PNA}),
which matches smoothly onto equation (\ref{eqn:PA}) in the limit $4n_0\gg 1$.
In addition we found that for the higher mass gaps including the possibility of
double resonances \cite{BahHax,Haxton} noticeably altered the results.
We assumed that for neutrinos produced at a radius $r_{prod}$ a fraction
\begin{equation}
  {1\over 2} \left( 1 - \sqrt{ 1 - (r_{res}/r_{prod})^2} \right)
\end{equation}
of them passed through two resonances.  This modifies the survival probability
by $P_{jump}\rightarrow 2P_{jump}(1-P_{jump})$ where $P_{jump}$ is the
coefficient of the $\cos 2\theta_m\cos 2\theta$ term in
(\ref{eqn:PNA}-\ref{eqn:PA}).

Finally to compute the survival probability in the SSM we averaged
(\ref{eqn:PNA}-\ref{eqn:PA}) over the $\nu$ production regions in the sun
\cite{TheBook}.

For each model and each $(\Delta m^2,\sin^2 2\theta)$ the survival
probabilities, along with the SSM $\nu_e$ spectra \cite{TheBook} were used
to compute $\nu_e$ and $\nu_{\mu}$ spectra at the Earth.  These spectra were
then convolved with simple detector efficiencies and cross sections
\cite{White,Gates,TheBook} to obtain predicted rates.
In addition we generated 100,000 ``flux variations" using a Monte-Carlo
procedure (as outlined in \cite{White}, based on earlier solar model
Monte-Carlos \cite{TheBook,BahHax}) to estimate the SSM errors in the
predictions.

The results of this stage then are a set of predicted average event rates in
the detectors (Homestake, Kamiokande, SAGE and GALLEX) with errors coming from
the SSM uncertainties.  These predicted rates $(p^{(i)}\pm \sigma_p^{(i)})$
were then compared with the experimental rates $(e^{(i)}\pm \sigma_e^{(i)})$ by
calculating the ratio
\begin{equation}
  r^{(i)}\pm \sigma_r^{(i)} =
  { p^{(i)}\pm\sigma_p^{(i)} \over e^{(i)}\pm\sigma_e^{(i)} }
\end{equation}
(combining errors in quadrature as usual) and computing $\chi^2$ for
a fit of the ratios to unity.
The averaged rates \cite{GALLEX,SAGE,Homestake,Kamiokande} used for the
fits are shown in table 1.

Finally, in order to estimate the effect of solar model uncertainties and our
fitting procedure, we compared the statistical fits described above, after
incorporating the results of our Monte Carlo over flux uncertainties with those
obtained before these uncertainties were incorporated. Note that the fits for
the data without standard solar model uncertainties were done using a standard
$\chi^2$ fitting procedure.  Thus the incorporation of solar model
uncertainties also changes the fitting procedure. Our results are
then displayed in figures 1-3.   As expected, the inclusion of the GALLEX
result is to reduce the allowed MSW phase space compared to that allowed for
SAGE alone. The inclusion of solar model uncertainties on the other hand
increases the region allowed at the 90\% confidence level for both MSW and
vacuum oscillations (this result should be contrasted to that obtained in
\cite{ken}).

As indicated above, spectral measurements and a search for possible day-night
variations can in principle further constrain the allowed regions displayed
here. Present spectral constraints from Kamiokande \cite{Kamiokande2} do not
yet provide further constraints on the MSW parameter space, but constraints
from day-night variations do. We display in the figures those regions of the
otherwise allowed MSW parameter space which are ruled out by these constraints
\cite{Kamiokande3}.

To conclude, the progress on the resolution of the solar neutrino problem has
been uneven.  While the data does favor a neutrino based solution,
unfortunately at present the results of the Gallium experiments are
sufficiently inconclusive so as to delay a definitive statement in this
regard.  Nevertheless, when all the existing time-averaged data is
incorporated, along with solar model uncertainties, a wide range of possible
neutrino masses and mixing angles, involving either matter or vacuum neutrino
oscillations, remains viable.  We may have to await a combination of high
statistics experiments and experiments with enhanced spectral sensitivity,
such as SNO \cite{SNO}, super-Kamiokande \cite{SKamio}, and perhaps Borexino
\cite{Borexino}, before a specific solution can be unambiguously determined.

We thank S. Petcov and D. Kennedy for useful discussions, and for describing
various details of their recent work on this subject to us. L.K. and E.G. also
thank the Aspen Center for Physics for hospitality during the course of this
work.

\clearpage

\noindent {\bf Figure Captions}

\vskip .2in

\noindent Figure 1: Region of mass-mixing angle space for neutrino
oscillation solutions allowed at the 90\% confidence level for the
combined Homestake-Kamiokande-SAGE data, (a) without, and (b) including
solar model uncertainties.  Within this region the portion
excluded by the day-night effect
in Kamiokande has been removed.

\vskip .1in

\noindent Figure 2: Region of mass-mixing angle space for neutrino
oscillation solutions allowed at the 90\% confidence level for the
combined Homestake-Kamiokande-GALLEX data, (a) without, and (b) including
solar model uncertainties.  Within this region the portion
excluded by the day-night effect
in Kamiokande has been removed.

\vskip .1in

\noindent Figure 3: Region of mass-mixing angle space for neutrino
oscillation solutions allowed at the 90\% confidence level for the
combined Homestake-Kamiokande-SAGE-GALLEX data, (a) without, and (b)
including solar model uncertainties.  Within this region the portion
excluded by the day-night effect
in Kamiokande has been removed.

\begin{table}
\begin{center}
\begin{tabular}{|l|c|}
\hline
Experiment & Rate  \\ \hline
Homestake  & $0.28 \pm 0.04$ \\
Kamiokande & $0.49 \pm 0.08$ \\
SAGE       & $0.44 \pm 0.21$ \\
GALLEX     & $0.63 \pm 0.16$ \\ \hline
\end{tabular}
\caption{The experimental rates, normalized to the standard solar model
predictions, used in the fits.}
\end{center}
\label{tab:expt-rates}
\end{table}

\end{document}